\documentclass[pdflatex,sn-mathphys-num]{sn-jnl}

\usepackage{graphicx}%
\usepackage{multirow}%
\usepackage{amsmath,amssymb,amsfonts}%
\usepackage{amsthm}%
\usepackage{mathrsfs}%
\usepackage[title]{appendix}%
\usepackage{xcolor}%
\usepackage{textcomp}%
\usepackage{manyfoot}%
\usepackage{booktabs}%
\usepackage{algorithm}%
\usepackage{algorithmicx}%
\usepackage{algpseudocode}%
\usepackage{listings}%
\usepackage{lineno}
\usepackage{lmodern}     
\usepackage{graphicx}    
\usepackage{caption}
\usepackage{subcaption}  
\usepackage{comment}     
\usepackage{enumitem}    
\usepackage{mathrsfs}
\usepackage{eucal}  
\usepackage[T1]{fontenc} 
\usepackage{url}
\usepackage{hyperref}
\usepackage{anyfontsize}
\usepackage{silence}
\begin{document}
\sloppy  
\title[Article Title]{Machine Learning-Driven Predictive Resource Management in Complex Science Workflows
}
%







\author*[1]{\fnm{Tasnuva} \sur{Chowdhury}}\email{tchowdhur@bnl.gov}

\author[1]{\fnm{Tadashi} \sur{Maeno}}
\author[5]{\fnm{Fatih Furkan} \sur{Akman}}
\author[3]{\fnm{Joseph} \sur{Boudreau}}
\author[1]{\fnm{Sankha} \sur{Dutta}}      
\author[4]{\fnm{Shengyu} \sur{Feng}}
\author[1]{\fnm{Adolfy} \sur{Hoisie}}
\author[1]{\fnm{Kuan-Chieh} \sur{Hsu}}
\author[3]{\fnm{Raees} \sur{Khan}}
\author[4]{\fnm{Jaehyung} \sur{Kim}}
\author[1]{\fnm{Ozgur O.} \sur{Kilic}}
\author[2]{\fnm{Scott} \sur{Klasky}}
\author[1]{\fnm{Alexei} \sur{Klimentov}}
\author[3]{\fnm{Tatiana} \sur{Korchuganova}}
\author[5]{\fnm{Verena Ingrid Martinez} \sur{Outschoorn}}
\author[1]{\fnm{Paul} \sur{Nilsson}}
\author[1]{\fnm{David K.} \sur{Park}}
\author[2]{\fnm{Norbert} \sur{Podhorszki}}
\author[4]{\fnm{Yihui} \sur{Ren}}
\author[5]{\fnm{John Rembrandt} \sur{Steele}}
\author[2]{\fnm{Fr\'ed\'eric} \sur{Suter}}
\author[1]{\fnm{Sairam Sri} \sur{Vatsavai}}
\author[1]{\fnm{Torre} \sur{Wenaus}}
\author[6]{\fnm{Wei} \sur{Yang}}
\author[4]{\fnm{Yiming} \sur{Yang}}
\author[1]{\fnm{Shinjae} \sur{Yoo}}

\affil*[1]{\orgname{Brookhaven National Laboratory}, \orgaddress{\city{Upton}, \state{NY}, \country{USA}}}
\affil[2]{\orgname{Oak Ridge National Laboratory}, \orgaddress{\city{Oak Ridge}, \state{TN}, \country{USA}}}
\affil[3]{\orgname{University of Pittsburgh}, \orgaddress{\city{Pittsburgh}, \state{PA}, \country{USA}}}
\affil[4]{\orgname{Carnegie Mellon University}, \orgaddress{\city{Pittsburgh}, \state{PA}, \country{USA}}}
\affil[5]{\orgname{University of Massachusetts}, \orgaddress{\city{Amherst}, \state{MA}, \country{USA}}}
\affil[6]{\orgname{SLAC National Accelerator Laboratory}, \orgaddress{\city{Menlo Park}, \state{CA}, \country{USA}}}


\abstract{The collaborative efforts of large communities in science experiments, often comprising thousands of global members, reflect a monumental commitment to exploration and discovery. Recently, advanced and complex data processing has gained increasing importance in science experiments. Data processing workflows typically consist of multiple intricate steps, and the precise specification of resource requirements is crucial for each step to allocate optimal resources for effective processing. Estimating resource requirements in advance is challenging due to a wide range of analysis scenarios, varying skill levels among community members, and the continuously increasing spectrum of computing options. One practical approach to mitigate these challenges involves initially processing a subset of each step to measure precise resource utilization from actual processing profiles before completing the entire step. While this two-staged approach enables processing on optimal resources for most of the workflow, it has drawbacks such as initial inaccuracies leading to potential failures and suboptimal resource usage, along with overhead from waiting for initial processing completion, which is critical for fast-turnaround analyses.\\
In this context, our study introduces a novel pipeline of machine learning models within a comprehensive workflow management system, the Production and Distributed Analysis (PanDA) system. These models employ advanced machine learning techniques to predict key resource requirements, overcoming challenges posed by limited upfront knowledge of characteristics at each step. Accurate forecasts of resource requirements enable informed and proactive decision-making in workflow management, enhancing the efficiency of handling diverse, complex workflows across heterogeneous resources.
}


\maketitle

\section{Introduction}
\label{sec1}

The collaborative efforts of large scientific communities, facilitated by sophisticated workflow management frameworks like the Production and Distributed Analysis (PanDA) system \cite{panda}, have demonstrated an enduring commitment to exploration and discovery. PanDA was originally developed to meet the extensive demands of the Large Hadron Collider (LHC) \cite{lhc} data processing ecosystem, specifically being successfully proven for the ATLAS experiment \cite{atlas} for almost two decades. It has effectively managed the complexities associated with data (re)processing, detector simulation, and physics analysis across a global network spanning approximately two hundred computing centers across forty countries. PanDA has adeptly supported a wide range of scientific applications and workflows, encompassing several billion hours of annual computing usage, thousands of scientists engaged in remote data analysis, and processing volumes exceeding the exabyte scale. PanDA's adaptability to emerging computing technologies has seamlessly integrated resources from the Worldwide LHC Computing Grid (WLCG) \cite{wlcg}, volunteer computing initiatives, high-performance computing (HPC) environments, leadership computing facilities (LCFs) \cite{alcf, olcf}, and commercial clouds. Its scalability and versatility position PanDA as an efficient solution for a variety of exabyte-scale scientific communities beyond high-energy physics, including applications at the Vera C. Rubin Observatory \cite{rubin}.\\

Recently, advanced and complex data processing has gained increasing importance in science experiments. Managing the escalating complexity of data processing workflows, presents challenges in precise resource allocation and optimal utilization. Estimating these requirements in advance is particularly challenging due to a wide range of analysis scenarios, varying expertise levels across communities, and the continuously increasing spectrum of computing options. To address these challenges, PanDA has employed a two-staged processing of workflow steps, where a subset of each step is initially processed to measure precise resource utilization from actual processing profiles before completing the entire step. This two-staged approach optimizes resource usage for most of the workflow, though it introduces initial inaccuracies that can lead to operational failures and suboptimal resource usage. Moreover, the associated overhead of waiting for initial processing completion poses critical constraints, particularly in contexts requiring rapid analysis turnaround times. Our research focuses on integrating advanced machine learning techniques within PanDA to predict and optimize key resource requirements, thereby eliminating the need for the two-staged approach. Our models overcome challenges stemming from limited upfront knowledge about step characteristics in complex workflows. They enable informed, proactive decision-making in workflow management and significantly enhance operational efficiency across diverse resources.\\

This paper begins by outlining the existing two-staged approach for estimating resource requirements, highlighting its issues. Subsequently, we introduce our machine learning-based solution and detail its implementation. This is followed by deployment in a test environment which mimics ATLAS production. Finally, we discuss the implications of our approach and its potential broader applications.

\section{PanDA Workflow Management System Overview}
\label{sec:panda-overview}

\begin{figure}[ht]
\centering
\includegraphics[width=\textwidth]{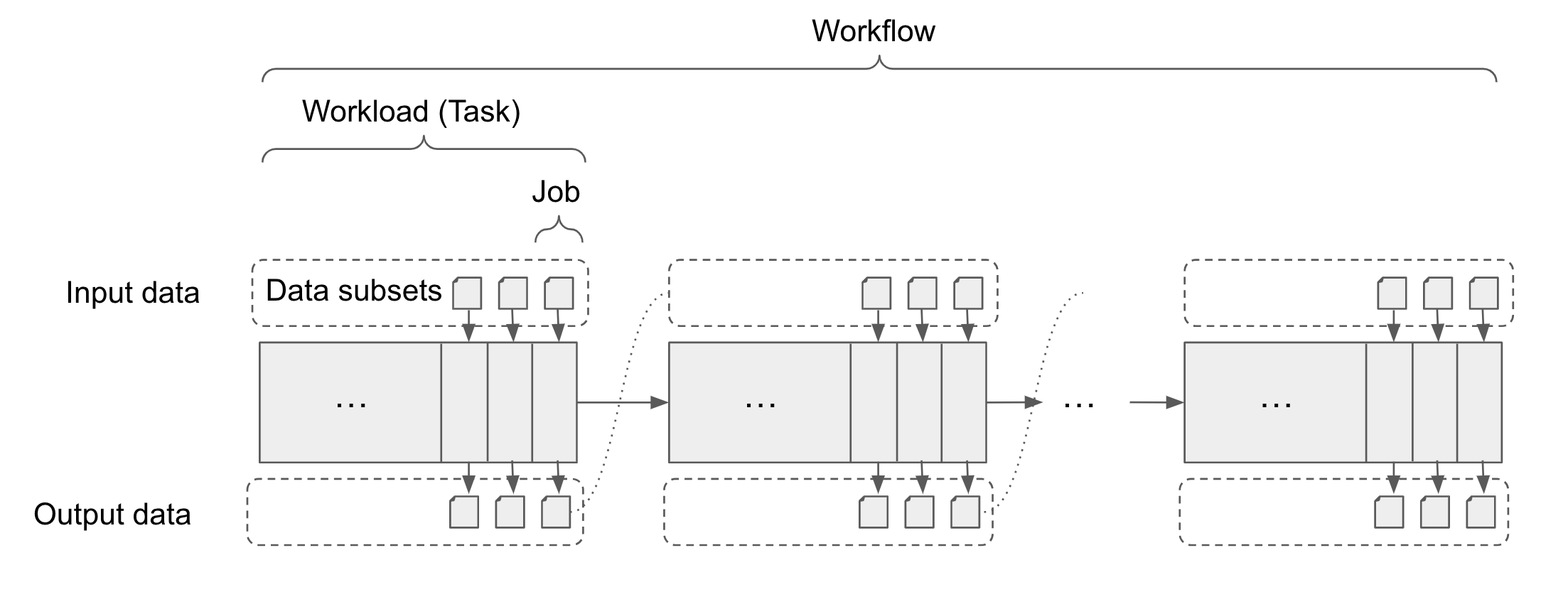}
\caption{Representation of a \textbf{basic workflow example} composed of a sequence of workloads that depend on the previous step \cite{panda}. Each workload is divided into jobs, which process an input data subset (e.g. a set of input files) and generate an output data subset (e.g. one output file). The output data of one workload is fed as input into the next workload in the sequence.}
\label{fig:workflow_hierarchy}
\end{figure}

Data processing workflows in PanDA are organized hierarchically into multiple steps with well-defined topological relationships. As illustrated in Figure. \ref{fig:workflow_hierarchy}, a workflow represents a scientific objective that is decomposed into intermediate stages, referred to as workloads. Examples of such objectives include searches for new particles, detector calibration in high-dimensional parameter spaces, and large-scale data processing for experimental collaborations.

Workloads within a workflow correspond to software applications that utilize computing resources to achieve one of these intermediate steps. In PanDA, workloads are represented by Tasks, which are further divided into Jobs. Jobs are the smallest executable entities and run independently in parallel, each processing a subset of input files to produce corresponding outputs. Resource requirements such as CPU cores, I/O intensity, memory usage, and wall time may differ across Tasks, reflecting the heterogeneity of the underlying computations.

The computing landscape available to PanDA encompasses a broad spectrum of resources, including the WLCG, volunteer computing initiatives, HPC environments, LCFs, commercial clouds, and emerging services such as superfacility API and Function as a Service platforms. Each environment offers distinct advantages, operational natures, and constraints.\\

Central to PanDA's resource allocation is the \textbf{Brokerage} component, a decision-making engine that assigns adapted resources to each Job based on its specific requirements. The Brokerage ensures efficient execution of workflows by matching Jobs to the most suitable resources available across the distributed infrastructure. This adaptability and scalability have enabled PanDA to support a wide range of scientific applications and workflows, making it an efficient solution for exabyte-scale data processing in high-energy physics and beyond.

\section{Challenges in Resource Estimation and Motivation for Machine Learning}
\label{sec:challenges-ml-motivation}

Despite the Brokerage's sophisticated matchmaking, a key obstacle is that resource requirements for each Job are not known in advance. This uncertainty arises from the diversity of analysis scenarios, the varying expertise levels among community members, and the rapidly evolving spectrum of computing options.\\

To address this, PanDA has traditionally employed a two-staged processing approach for each Task. Initially, a subset of Jobs, known as \textbf{Scout Jobs}, is executed to empirically measure resource utilization. The results from these Scout Jobs inform the resource allocation for the remaining Jobs in the Task, aiming to optimize resource usage and minimize failures.\\

However, this two-staged approach presents several notable drawbacks:
\begin{itemize}
    \item \textbf{Resource Misallocation:} Initial uncertainties about resource requirements can lead to misallocation for Scout Jobs, resulting in failures or inefficient usage.
    \item \textbf{Operational Overhead:} The necessity to wait for Scout Job completion introduces significant overhead, especially problematic for analyses requiring rapid turnaround.
    \item \textbf{Limited Scalability:} As workflows grow in complexity and scale, the inefficiencies and delays associated with Scout Jobs become increasingly pronounced.
\end{itemize}
\color{black}

\begin{figure}[htbp]
    \centering
    \includegraphics[width=0.7\textwidth]{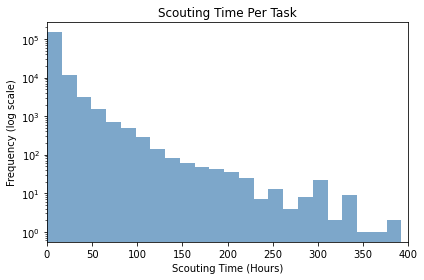}
    \caption{Completion of scout jobs per Task.}
    \label{fig:scout-wait-times}
\end{figure}

The histogram in Figure~\ref{fig:scout-wait-times} reveals that while the majority of Scout Jobs are completed within a relatively short time frame, a non-negligible tail exists: out of hundreds of thousands, a few thousand jobs require more than 150 hours to complete. These outliers can significantly impact the overall turnaround time for large-scale analyses.
\newline
To overcome these limitations, our objective is to develop machine learning models capable of accurately predicting resource requirements at the outset, thereby eliminating the need for Scout Jobs and their associated inefficiencies. The key challenges in this approach include the limited availability of task-specific information early in the workflow and the necessity to leverage a restricted feature set for initial predictions.\\

Our research aims to achieve the following:
\begin{itemize}
    \item  Optimize resource management and task completion times by using precise predictive models for memory, CPU time, and I/O, walltime reducing the need for Scout Jobs.
    \item Establish a feedback loop within the machine learning pipeline, where predictions are continuously refined based on actual Job performance, iteratively improving estimation accuracy.
    \item Evaluate and compare the performance of our machine learning models against existing methods, demonstrating potential enhancements in operational efficiency within real-world scientific experiments.
\end{itemize}

\section{Machine Learning-Based Predictive Resource Management}
\label{sec:ml-prediction}
We now present our machine learning-based solution for predictive resource management in the PanDA workflow system. This section outlines the dataset utilized, available features for early-stage prediction, critical target variables for efficient workflow execution, data preprocessing for discretization into distinct classes, the design of our classification-based prediction pipeline, and the evaluation methodology employed.

\subsection{Data, Features, Target Variables, and Preprocessing}
\label{sec:data-analysis}

Our study leverages four years of operational data from the PanDA Workflow Management System, comprising 4M successful task executions. The dataset includes a diverse set of features available at the time of task submission, enabling resource prediction even before Scout Job initialization. \\

The dataset contains features like \texttt{PROCESSINGTYPE} specifying the processing such as derivation and simulation, \texttt{FRAMEWORK} indicates software environment  such as Athena and AthSimulation, and NCORE details the parallelism configuration (single or multi-core usage). Additionally, workload scale is characterized by \texttt{NINPUT} (the total number of unique input file sets), \texttt{NFILES} (the number of files within those file sets), and \texttt{NEVENTS} (the total number of events to be processed). These features are selected as they are reliably available prior to Scout Job execution and have demonstrated predictive value for downstream resource utilization.\\

Our predictive framework is designed to accurately estimate four key resource metrics from actual job profiles, providing robust targets for efficient workflow management. For each metric---RAM Allocation, CPU Time, I/O Intensity, and Walltime---we compute representative values based on execution profiles from real jobs, using established computational procedures. This ensures that resource predictions are firmly anchored in empirical job performance and are appropriately tailored to the requirements of subsequent workflow steps. The definitions of our four target variables are given as follows.\\

\begin{itemize}
    \item \textbf{RAM Allocation (\texttt{RAMCOUNT}):} Per-core memory requirement for each job is computed as
    \begin{equation}
        \text{ramCount} = \max\left(\frac{\text{maxPSS} - \text{baseRamCount}}{\text{coreCount}} \times \text{margin}, \text{minRamCount}\right)
    \end{equation}
    where \texttt{maxPSS} is the peak resident set size, \texttt{baseRamCount} is a baseline offset, \texttt{coreCount} is the number of CPU cores, and \texttt{margin} (default 10) and \texttt{minRamCount} are configurable safety parameters. The 75th percentile of scout job \texttt{ramCount} is used to estimate expected memory usage for subsequent jobs.\\
    
    \item \textbf{CPU Time (\texttt{CPUTIME}):} The expected CPU time per event is estimated using

    \begin{align}
    \text{cpuTime} =\ & 
    \frac{
        \max\big(0,\, \text{endTime} - \text{startTime} - \text{baseTime}\big) \times \text{corePower}
    }{ 
        \text{nEvents}
    }
    \notag \\
    & \times\, \text{coreCount} \times \text{cpuEfficiency} \times 1.5
    \end{align}

    where \texttt{corePower} is the HS06 rating of the resource, \texttt{cpuEfficiency} defaults to 90\%, and \texttt{nEvents} is the total number of events. To ensure reliability, we use the 95th percentile of \texttt{cpuTime} from scout jobs with at least $10 \times$ \texttt{coreCount} events or jobs longer than 6 hours.\\
    
    \item \textbf{I/O Intensity (\texttt{IOINTENSITY}):} This metric quantifies the data throughput as the ratio of total input and output data size to execution time, providing insight into the job's impact on storage and network resources.\\

    \item \textbf{Walltime (\texttt{WALLTIME}):} The estimated walltime for a job is
    \begin{equation}
        \text{walltime} = \frac{\text{cpuTime} \times \text{nEvents}}{C \times P \times \text{cpuEfficiency}} + \text{baseTime}
    \end{equation}
    where $nEvents$ is the number of events, $C$ and $P$ are configuration parameters, and \texttt{cpuEfficiency} is as above. The estimated \texttt{walltime} must be between predefined \texttt{mintime} and \texttt{maxtime} values at the queue.\\
\end{itemize}

\begin{figure}[htbp]
    \centering
    \includegraphics[width=.9\textwidth]{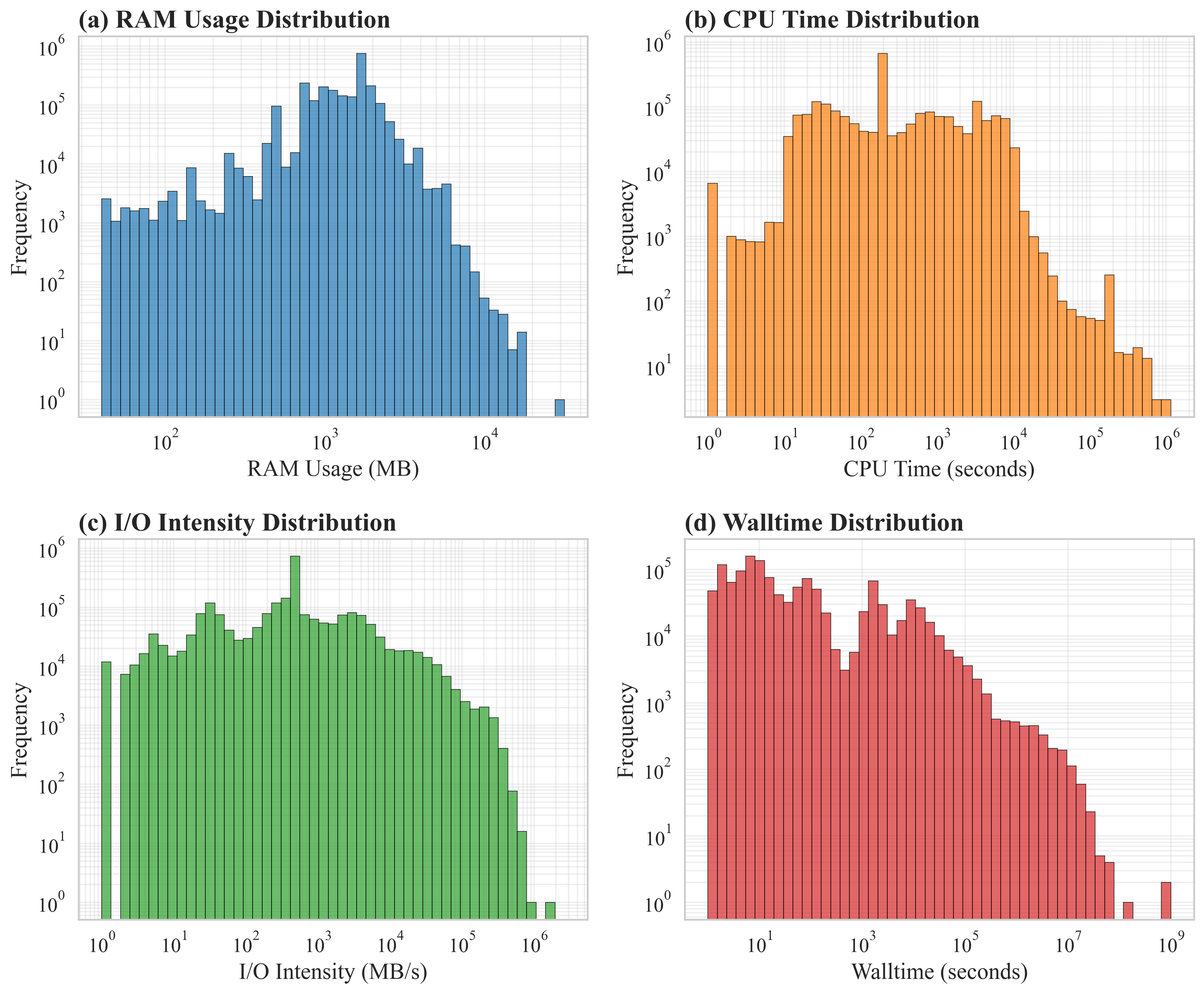}
    \caption{Distributions of key resource metrics across all tasks: 
    (a) RAM usage, (b) CPU time, (c) I/O intensity, and (d) Walltime. Both axes are log-scaled to highlight the heavy-tailed nature of these resource requirements.}
    \label{fig:target_variables}
\end{figure}

\begin{figure}[htbp]
    \centering
    \includegraphics[width=.9\textwidth]{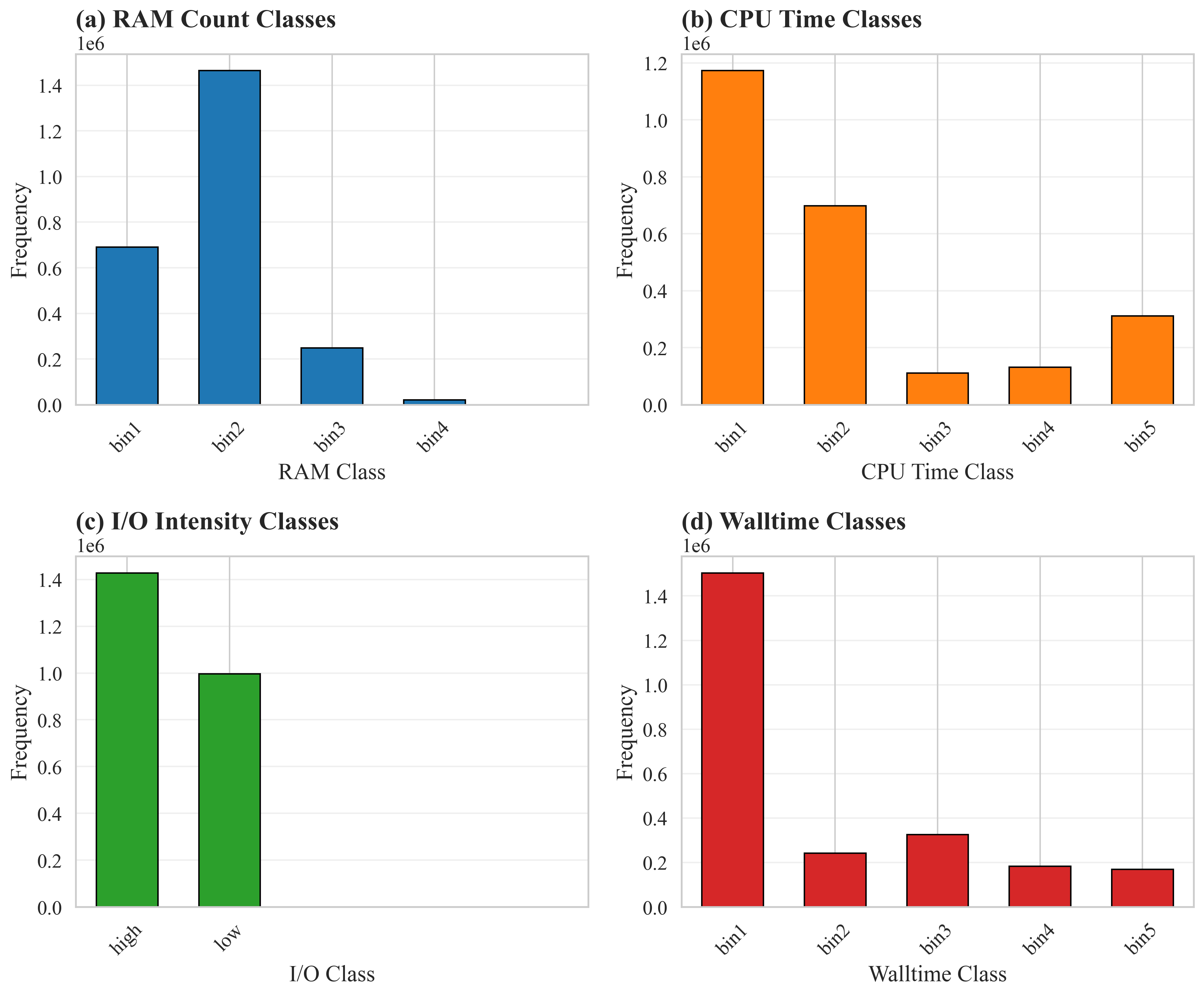}
    \caption{Distributions of discretized resource classes: (a) RAM Count, (b) CPU Time, (c) I/O Intensity and (d) Walltime. The class-based representation enables robust multi-class classification and aligns with operational resource allocation strategies.}
    \label{fig:class_distributions}
\end{figure}

The raw distributions of the four primary target variables are highly skewed and span several orders of magnitude, as shown in Figure~\ref{fig:target_variables}. Such variability poses challenges for direct regression and robust resource allocation. To address this, we discretize each target variable into a small number of well-defined classes, transforming the prediction task into a multi-class classification problem. For example, RAM Count, CPU Time, and Walltime are binned into categories labeled as \emph{Bin 1}, \emph{Bin 2}, \emph{Bin 3}, and so forth, according to quantiles or operationally meaningful thresholds, while I/O Intensity is stratified into \emph{High} and \emph{Low} classes. This transformation, illustrated in Figure~\ref{fig:class_distributions}, not only mitigates the impact of extreme values but also aligns predictions with the decision-making requirements of the PanDA brokerage system, which operates on discrete resource allocation tiers. By converting continuous resource metrics into classes, our models are better equipped to provide actionable and reliable guidance for workflow scheduling and resource provisioning. Table~\ref{tab:model_mapping} provides an overview of each predictive model, indicating the corresponding target variable and the number of classes used for classification.\\

\begin{table}[htbp]
    \centering
    \begin{tabular}{@{}llc@{}}
        \toprule
        \textbf{Model} & \textbf{Predicted Target}        & \textbf{Number of Classes} \\
        \midrule
        Model 1 & RAM Classification (\texttt{RAMCOUNT})    & 4  \\
        Model 2 & CPU Time Classification (\texttt{CPUTIME})           & 5  \\
        Model 3 & I/O Intensity Classification (\texttt{IOINTENSITY})  & 2 \\
        Model 4 & Walltime Classification (\texttt{WALLTIME})  & 5 \\
        \bottomrule
    \end{tabular}
    \caption{Classification models and their associated target variables.\\}
    \label{tab:model_mapping}
\end{table}

\subsection{Model Architectures: Empirical Baseline and Deep Learning}

A robust baseline for resource requirement classification was established using gradient-boosted decision trees implemented through the XGBoost framework~\cite{chen2016xgboost}. Categorical workflow metadata features were label-encoded, and hyperparameters such as maximum tree depth, learning rate, and regularization parameters were tuned via grid search with cross-validation. This well-established approach provides an empirical benchmark due to XGBoost’s proven efficiency and scalability in structured data applications.\\

Building upon this baseline, a deep learning model was designed to capture complex, nonlinear dependencies across heterogeneous workflow metadata. The model employs adaptive embedding layers for categorical variables, which learn dense vector representations that compactly encode latent feature relationships beyond what fixed encodings provide. The embedding dimension for a categorical feature is defined as

\[
\text{embed\_dim} = \min(32, \lfloor \log_2(v) \rfloor + 1),
\]

where \(v\) denotes the feature’s cardinality. Numerical features are concatenated directly with the embedded categorical representations, forming a unified input vector processed through a sequence of fully connected layers. Batch normalization~\cite{ioffe2015batch} is applied after each layer to stabilize training, while dropout and \(L_2\) regularization are employed to control overfitting. Final architectural parameters and optimization details are described in Section~\ref{sec:model_tuning}. \\

Due to the computational expense of neural network training, hyperparameter exploration used a random search approach to efficiently traverse the high-dimensional configuration space~\cite{bergstra2012random}. The resulting architecture serves as the foundation of the proposed predictive framework for resource requirement estimation in large-scale scientific workflows.

\subsection{Model Training and Hyperparameter Tuning}
\label{sec:model_tuning}

The dataset comprised approximately 3.4M tasks, representing 85\% of the initial data. This dataset was further partitioned into training/validation (85\%) and test (15\%) subsets using stratified sampling to preserve the class distribution across splits~\cite{kohavi1995study}. Hyperparameter optimization was performed using three-fold cross-validation combined with a randomized search strategy~\cite{bergstra2012random}. The search identified an optimal network configuration consisting of three fully connected layers containing 256, 128, and 64 units, respectively, with progressive dropout rates set to 40\%, 30\%, and 30\%. An \(L_2\) regularization coefficient of \(\lambda = 10^{-4}\) was applied to all dense layers’ kernels, and the learning rate was fixed at \(5 \times 10^{-5}\). Using these hyperparameters, the final model was retrained on the entire 3.4M tasks and evaluated on 0.6M unseen data.\\

To account for class imbalance in workflow classifications, class weights inversely proportional to class frequencies were incorporated into the loss function~\cite{he2009learning}, ensuring balanced learning across categories. Early stopping with a patience of four epochs based on validation accuracy was applied, and the best-performing weights were restored to prevent overfitting~\cite{ioffe2015batch}. \\

Training utilized the Adam optimizer~\cite{kingma2014adam} due to its adaptive learning rate properties and convergence stability. Sparse categorical cross-entropy was used for multi-class resource classification, and binary cross-entropy for binary I/O predictions.\\

This combination of stratified sampling, randomized search tuning, class weighting, early stopping, and adaptive optimization ensures both high predictive accuracy and strong generalization capability. Benchmarking against the XGBoost baseline was conducted using identical data partitions for a fair comparative analysis, as presented in Section~\ref{eval_sec}.

\color{black}
\subsection{Evaluation Metrics and Model Performance}
\label{eval_sec}
To comprehensively assess model performance, a combination of standard metrics suitable for both binary and multiclass classification problems was employed. Overall \textbf{accuracy} quantifies the proportion of correctly classified instances and provides a basic indicator of prediction reliability. However, since the resource classification tasks exhibit substantial class imbalance,  we present additional metrics that offer a more granular evaluation~\cite{cambridge_ml_evaluation, nature_metrics, arxiv2404_07661}.\\

Both the deep learning (DL) models and the XGBoost baseline were evaluated on the held-out test set to determine their predictive capabilities for resource requirement classification, as summarized in Table~\ref{tab:model_comparison}. While XGBoost—with label-encoded categorical features—provides a robust empirical benchmark with competitive accuracy and precision, the DL models demonstrate similar or slightly higher performance. Furthermore, the DL framework offers greater extensibility by incorporating additional features and capturing richer representations for future enhancements.\\


\begin{table}[h]
\centering
\begin{tabular}{@{}lcccc@{}}
\toprule
\textbf{Model} & \textbf{Metric} & \textbf{XGBoost} & \textbf{Deep Learning} \\
\midrule
\multirow{4}{*}{Model 1 } 
    & Accuracy         & 0.80 & 0.88 \\
    & Macro F1         & 0.60 & 0.73 \\
    & ROC-AUC$^{*}$    & 0.97 & 0.98 \\
    & PR-AUC$^{*}$   & 0.92 & 0.96 \\
\midrule
\multirow{4}{*}{Model 2 } 
    & Accuracy         & 0.85 & 0.86 \\
    & Macro F1         & 0.72 & 0.73 \\
    & ROC-AUC$^{*}$    & 0.98 & 0.99 \\
    & PR-AUC$^{*}$   & 0.95 & 0.95 \\
\midrule
\multirow{4}{*}{Model 3 } 
    & Accuracy         & 0.88 & 0.94 \\
    & Macro F1         & 0.88 & 0.94 \\
    & ROC-AUC     & 0.96 & 0.99 \\
    & PR-AUC    & 0.96 & 0.98 \\
\midrule
\multirow{4}{*}{Model 4 } 
    & Accuracy         & 0.83 & 0.91 \\
    & Macro F1         & 0.67 & 0.80 \\
    & ROC-AUC$^{*}$    & 0.98 & 0.99 \\
    & PR-AUC$^{*}$   & 0.93 & 0.97 \\
\bottomrule
\multicolumn{4}{l}{\footnotesize $^{*}$Micro-average AUC for multiclass models.} \\
\end{tabular}
\caption{Summary of Performance Comparison of XGBoost and Deep Learning Models}
\label{tab:model_comparison}
\end{table}

\color{black}

\begin{table}[ht]
\centering
\begin{tabular}{llccc}
\textbf{Model} & \textbf{Class} & \textbf{Precision} & \textbf{Recall} & \textbf{F1-score} \\
\hline
Model 1 & bin1  & 0.74 & 0.95 & 0.83 \\
        & bin2  & 0.99 & 0.86 & 0.92 \\
        & bin3  & 0.52 & 0.90 & 0.66 \\
        & bin4  & 0.35 & 0.92 & 0.50 \\
\hline
Model 2 & bin1  & 0.97 & 0.92 & 0.94 \\
        & bin2  & 0.90 & 0.91 & 0.91 \\
        & bin3  & 0.46 & 0.60 & 0.52 \\
        & bin4  & 0.40 & 0.69 & 0.51 \\
        & bin5  & 0.90 & 0.64 & 0.75 \\
\hline
Model 3 & low   & 0.96 & 0.94 & 0.95 \\
        & high  & 0.92 & 0.95 & 0.94 \\
\hline
Model 4 & bin1  & 1.00 & 0.93 & 0.96 \\
        & bin2  & 0.50 & 0.91 & 0.64 \\
        & bin3  & 0.80 & 0.77 & 0.78 \\
        & bin4  & 0.75 & 0.87 & 0.80 \\
        & bin5  & 0.77 & 0.85 & 0.81 \\
\hline
\end{tabular}
\caption{Precision, recall, and f1-score for each class and model.}
\label{tab:prf_per_class}
\end{table}

Table~\ref{tab:prf_per_class} details the \textbf{precision}, \textbf{recall}, and \textbf{F1-score} for each individual class of every model ~\cite{cambridge_ml_evaluation}. These classwise metrics, together with their macro and micro aggregations summarized in Table~\ref{tab:model_comparison}, help elucidate both overall trends and individual class performance. For multiclass tasks, metrics are reported for each resource consumption bin, while the binary I/O intensity model reports for both "low" and "high" classes. \\

\begin{figure}[htbp]
    \centering
    \begin{subfigure}{0.47\textwidth}
        \centering
        \includegraphics[width=\linewidth]{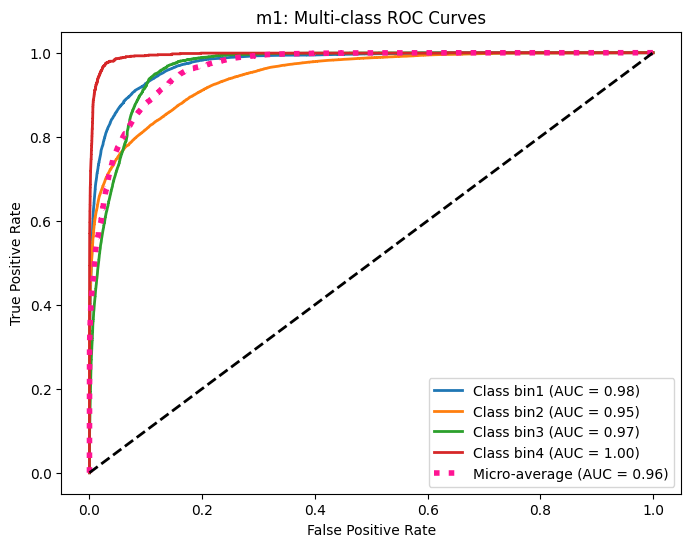}
        \caption{Model 1: RAM Classification}
        \label{fig:roc_m1}
    \end{subfigure}
    \hfill
    \begin{subfigure}{0.47\textwidth}
        \centering
        \includegraphics[width=\linewidth]{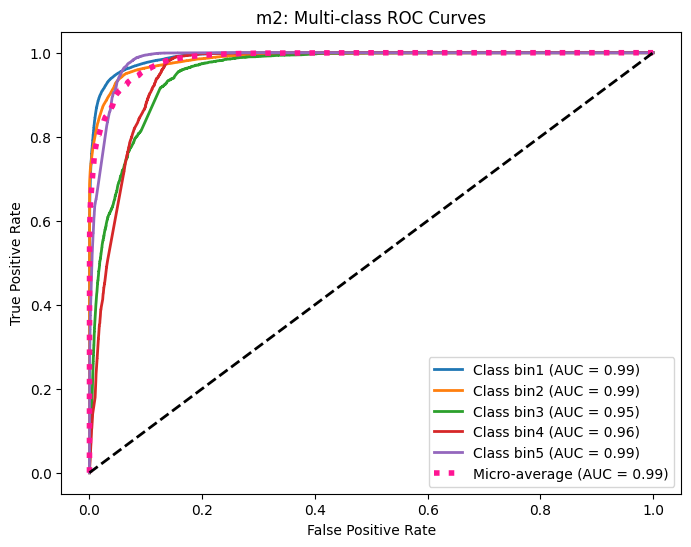}
        \caption{Model 2: CPU Time Classification}
        \label{fig:roc_m2}
    \end{subfigure}
    \\[1em]
    \begin{subfigure}{0.47\textwidth}
        \centering
        \includegraphics[width=\linewidth]{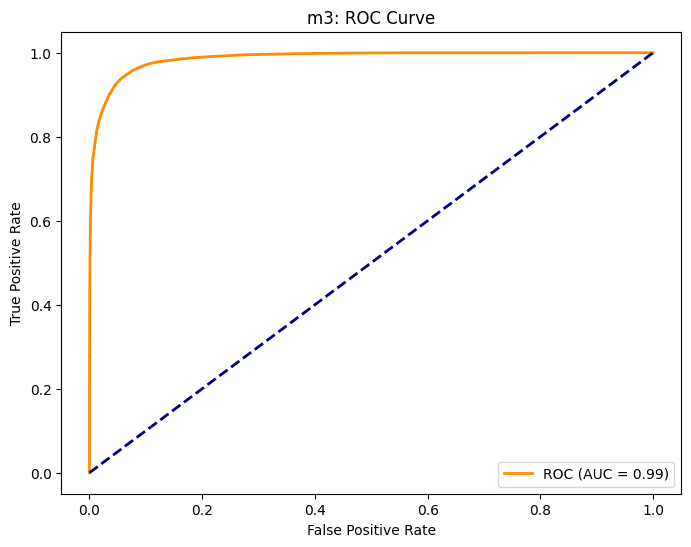}
        \caption{Model 3: I/O Intensity (Binary)}
        \label{fig:roc_m3}
    \end{subfigure}
    \hfill
    \begin{subfigure}{0.47\textwidth}
        \centering
        \includegraphics[width=\linewidth]{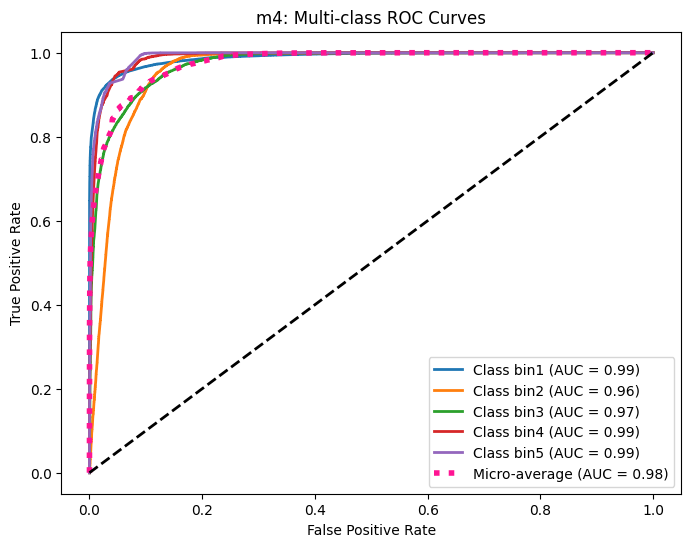}
        \caption{Model 4: Walltime Classification}
        \label{fig:roc_m4}
    \end{subfigure}
    \caption{Receiver Operating Characteristic (ROC) curves for all models. (a) Model 1: Multi-class ROC curves for RAM classification. (b) Model 2: Multi-class ROC curves for CPU time classification. (c) Model 3: Binary ROC curve for I/O intensity classification. (d) Model 4: Multi-class ROC curves for walltime classification. Micro-average AUC is shown for multiclass models.}
    \label{fig:all_rocs}
\end{figure}

To further evaluate discriminative capability, we use the area under the receiver operating characteristic curve (\textbf{ROC-AUC}), with micro-averaged values provided in Table~\ref{tab:model_comparison}. In the multiclass context, this gives a global view of the model's ability to separate all classes across threshold settings. Figure~\ref{fig:all_rocs} shows the ROC curves for all models, visually demonstrating the quality of classification boundaries throughout the resource estimation pipeline. For binary classification tasks, including the I/O intensity model, the ROC curve provides insight into the model’s performance independent of class distribution~\cite{arxiv2404_07661}. \\

\textbf{Model 1} (RAM classification) attains 88\% accuracy and macro F1 of 0.73, with a high ROC-AUC of 0.98. Notably, classwise results in Table~\ref{tab:prf_per_class} indicate the model achieves especially strong recall in bin1 (0.95) and bin4 (0.92). 
\textbf{Model 2} (CPU time classification) matches this overall performance, with 86\% accuracy, macro F1 of 0.73, and ROC-AUC of 0.99; the lowest per-class F1 scores (bin3: 0.52, bin4: 0.51) are due to class rarity but remain operationally useful.\\

\textbf{Model 3}, a binary classifier for I/O intensity, obtains 94\% accuracy and macro F1 of 0.94, along with a near-perfect ROC-AUC of 0.99. Both “low” and “high” I/O intensity classes are identified robustly, as confirmed by the ROC curve in Figure~\ref{fig:all_rocs}c. For \textbf{Model 4} (walltime classification), accuracy reaches 91\% with macro F1 of 0.80 and ROC-AUC of 0.99, indicating uniformly high performance across the five prediction bins.\\

All metrics are evaluated on held-out test sets comprising 15\% of the data, and model selection is guided by validation set performance and class-weighted loss to mitigate data imbalance~\cite{nature_metrics}. As shown in Figure~\ref{fig:all_rocs}, the ROC curves further demonstrate the ability of each model to distinguish target classes across threshold settings, with consistently strong micro-average AUC values underscoring high discriminative effectiveness.\\

\begin{figure}[htbp]
\centering
\hspace*{-5pt}
\includegraphics[width=1.05\textwidth]{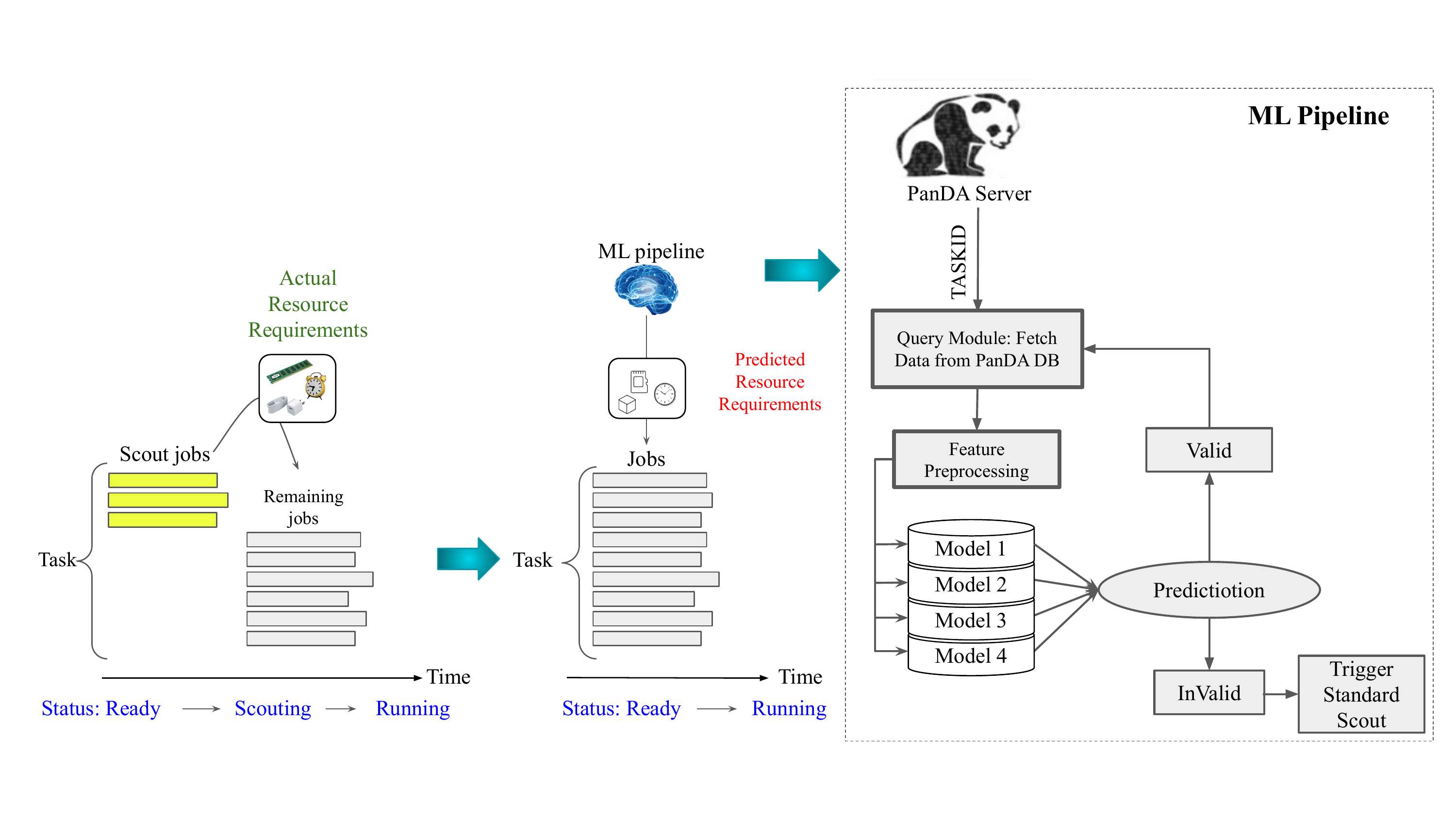} 
\caption{ML prediction service for the PanDA system.}
\label{fig:pipeline-architecture}
\end{figure}

\section{System Integration and Workflow Impact}
\label{sec:integration-impact}
We have developed a custom service-oriented architecture for integrating machine learning–driven resource prediction into the PanDA workload management system. 
The architecture is currently being evaluated in a dedicated test environment that closely replicates production conditions, enabling thorough validation with realistic scientific workloads. This test environment is essentially a reproduction of the production setup where real-time data used in production by the traditional Scout model is concurrently utilized by the machine learning pipeline for predictions. As depicted in Figure~\ref{fig:pipeline-architecture}, while the traditional Scout model continuously operates in the production environment, the machine learning pipeline ingests the same live data within the test environment. This setup enables direct comparison and validation of the ML pipeline's predictive performance and reliability against the benchmark traditional system under identical workload conditions, ensuring rigorous and practical evaluation prior to full-scale deployment.
\color{black}
As shown in Figure~\ref{fig:pipeline-architecture}, machine learning components are exposed as microservices within the workflow, deployed in containers to ensure consistent and reproducible execution. This approach facilitates automated resource prediction, precise scheduling, and streamlined workflow progression, with results from the test environment informing future production deployment.

\subsection{Integrated Pipeline Evaluation for PanDA Brokerage}
\label{subsec:pipeline_evaluation}
To quantify the effectiveness of our integrated solution, the PanDA brokerage system employs four predictive deep learning models that operate concurrently, providing real-time resource allocation estimates for incoming tasks. Assessing the overall system performance requires joint evaluation metrics that capture the collective contributions and interactions of these models within the broader workflow. Such an approach enables a holistic understanding of how the predictive ensemble drives improvements in scheduling accuracy, resource utilization, and workflow efficiency~\cite{cambridge_ml_evaluation}. This integrated evaluation informs future refinements and will guide deployment in the full production environment.

\subsubsection*{Pipeline Accuracy Calculation}

We assess end-to-end performance using the \emph{exact-match} (subset) accuracy metric, a standard in multi-label and multi-output classification for demanding applications such as high-performance computing and large-scale workload management~\cite{mlc_hpc_workload, mlc_comp_study}. Subset accuracy measures the proportion of tasks for which \emph{all} model outputs simultaneously match their ground truth, providing a stringent indicator of overall system reliability. Formally, pipeline accuracy $A$ is defined as:
\[
A = \frac{1}{N} \sum_{i=1}^N \mathbb{I} \left[ y_{i,1} = \hat{y}_{i,1} \land y_{i,2} = \hat{y}_{i,2} \land y_{i,3} = \hat{y}_{i,3} \land y_{i,4} = \hat{y}_{i,4} \right]
\]
where $N$ is the number of samples, $y_{i,j}$ and $\hat{y}_{i,j}$ are the true and predicted labels for task $j$ on sample $i$, and $\mathbb{I}(\cdot)$ is the indicator function.\\

To provide a broader view of system usability in production, we also report \emph{partial accuracy} metrics—the proportion of samples for which at least $k$ out of four predictions are correct. These metrics are useful when partial success is valuable for fail-over or fallback strategies, or when system degradation needs to be quantified. Table~\ref{tab:pipeline-accuracy} summarizes the strict and partial accuracy results on our test environment for about \textasciitilde100,000 tasks, reported as percentages.

\begin{table}[h]
    \centering
    \begin{tabular}{l c}
        \hline
        \textbf{Metric} & \textbf{Accuracy (\%)} \\
        \hline
        At least 1/4 correct & 99.8 \\
        At least 2/4 correct & 97.9 \\
        At least 3/4 correct & 89.1 \\
        At least 4/4 correct & 57.4  \\
        \hline
    \end{tabular}
    \caption{Pipeline accuracy and partial accuracy metrics on the test set (\%).}
    \label{tab:pipeline-accuracy}
\end{table}

\subsubsection*{Average Pipeline Accuracy}

Beyond strict exact-match accuracy, we compute the \emph{average pipeline accuracy}, where each of the four model outputs is considered as an equally weighted contributor to the overall performance. This metric, defined as the mean of the individual model accuracies, provides a more lenient and interpretable summary of typical model behavior~\cite{mlc_comp_study, pipeline_comp_framework}.\\

Formally, let the accuracy of model $j$ be denoted by $A_j$. The average pipeline accuracy $\bar{A}$ is then defined as
\[
\bar{A} = \frac{1}{4} \sum_{j=1}^{4} A_j
\]
where each $A_j$ is measured on approximately 100{,}000 tasks.
For our pipeline, this results in
\[
\bar{A} = \frac{80.5 + 85.8 + 93.9 + 83.9}{4} = 86.03\%
\]
Thus, the average pipeline accuracy reflects the expected probability that an individual model prediction within the pipeline is correct, under the assumption that all four resource prediction tasks are equally important.\\

\subsubsection*{Operational Metrics}

For practical deployment and robust evaluation of the pipeline, we recommend tracking several summary metrics that capture both strict and relaxed perspectives on performance:\\

\begin{itemize}
  \item \textbf{Joint Prediction Success Rate}: The proportion of cases where all four tasks are predicted correctly (\textbf{57.4\%}). This offers a strict measure of overall pipeline accuracy.
  \item \textbf{Average Model Accuracy}: The mean accuracy across the four individual models (\textbf{86.03\%}), providing a view of typical per-task correctness.
  \item \textbf{At-Least-One-Correct Rate}: The percentage of cases where at least one model produces a correct prediction (\textbf{99.8\%}), relevant for scenarios allowing partial credit or fallback.\\
\end{itemize}

Taken together, these metrics provide a transparent and comprehensive assessment of the pipeline, characterizing both all-or-nothing success and relaxed, real-world fallback behavior. This framework supplies a robust and interpretable operational baseline under the assumption of model independence.

\subsection{Performance Benchmark and Workflow Impact}

The integration of machine learning into the PanDA workflow system delivers transformative performance improvements. With sub-second inference latency, the ML pipeline achieves an orders-of-magnitude speedup over the traditional Scout-based approach, which required hours of task execution to estimate resource needs. Table~\ref{tab:performance-comparison} summarizes the key performance metrics:\\

\begin{table}[htbp]
\centering
\begin{tabular}{lcc}
\toprule
\textbf{Metric} & \textbf{ML Pipeline} & \textbf{Traditional Scout} \\
\midrule
Average Prediction Time (per task) & $<$ 1 second & $\approx$ 7 hours \\
\bottomrule
\end{tabular}
\caption{Performance comparison between the ML-driven pipeline and traditional Scout-based approach.}
\label{tab:performance-comparison}
\end{table}

This leap in performance fundamentally transforms workflow scheduling and execution in PanDA, enabling:\\

\begin{itemize}
    \item \textbf{Parallel prediction capability}: All four classification models generate predictions simultaneously within a single inference pass, ensuring efficient, per-task decision making at scale.
    
    \item \textbf{Dynamic, just-in-time resource allocation}: Real-time predictions facilitate prompt provisioning of CPU, memory, and I/O capacities, optimizing system responsiveness.
    
    \item \textbf{Enhanced scheduling flexibility}: Near-instantaneous predictions allow for opportunistic job scheduling on transient or heterogeneous resources.
    
    \item \textbf{Support for advanced workflow patterns}: The ML-driven system enables sophisticated workflow dynamics, including real-time resource negotiation and dynamic parameter tuning—both of which were previously impractical using static estimation methods.\\
\end{itemize}

The ML-integrated pipeline has been successfully tested in test environment, executing thousands of tasks with high service availability and reliability across diverse computing platforms. This signals a paradigm shift from static, latency-bound estimation mechanisms to a robust, agile, and learning-driven resource management framework that is scalable and well-suited for the future demands of distributed scientific computing.\\

\section{Conclusion}
\label{sec:conclusion}

This work introduces a comprehensive machine learning pipeline for dynamic resource prediction and allocation within the PanDA brokerage system. The pipeline demonstrates robust predictive capabilities across key resource types—including memory usage, CPU time, I/O intensity, and walltime—successfully distinguishing between diverse resource demand profiles.\\

In this study, we developed and rigorously evaluated machine learning models for resource requirement prediction in scientific workflows. Gradient-boosted decision trees (XGBoost) served as a strong empirical baseline, while deep neural networks with learned embeddings for categorical metadata achieved enhanced predictive accuracy and generalization. The use of adaptive embeddings and extensive regularization allowed the deep models to capture hidden relationships in workflow metadata, outperforming the baseline in both predictive stability and scalability. Comprehensive evaluation confirmed that the deep learning framework is well-suited for production integration, providing a reliable foundation for operational deployment within large-scale computing systems.\\

This work represents a significant advancement in intelligent resource management for large-scale scientific workflows. By providing accurate, early-stage predictions, our models help streamline resource allocation, reduce failures, and improve overall system throughput. Future work will focus on enhancing model adaptability across diverse workflow types and incorporating feedback mechanisms for continual model refinement.\\

Evaluation of the integrated solution confirms strong end-to-end performance, with consistent alignment of predictions across all models for a substantial fraction of tasks. Individual components maintain high accuracy overall, and the system commonly provides at least one correct prediction per task, ensuring practical reliability.\\

\color{black}
Deploying this pipeline in PanDA will result in significant operational gains, transforming resource assignment from a two-stage process with a delay into a seamless, single-step operation without latency. This advancement supports reliable, high-throughput computing and more efficient use of large-scale resources, marking an important step forward for intelligent workload management.\\

Looking forward, several research directions could further enhance this approach. Integrating clustering of specific task attributes and incorporating domain-specific knowledge could enhance the adaptability and precision of resource prediction. Establishing feedback-driven model refinement could continuously improve performance through systematic retraining based on operational outcomes. Additionally, exploring transfer learning across diverse scientific workflows could make these models more versatile, while reducing the need for extensive retraining. Pursuing these avenues will help develop more adaptive, efficient, and broadly applicable solutions for intelligent resource management in data-intensive computing environments.\\

\section*{Acknowledgments}

This material is based on work supported by the U.S. Department of Energy, Office of Science, Office of Advanced Scientific Computing Research under Award Number DE-SC-0012704 (REDWOOD project). This work was done in collaboration with the distributed computing project within the ATLAS Collaboration. We thank our ATLAS colleagues for their support, particularly the ATLAS Distributed Computing team's contributions. We would also like to express
our deepest gratitude to Prof. Kaushik De at the University of Texas at Arlington.

\bibliography{sn-bibliography}

\end{document}